\documentclass[11pt]{article}
\usepackage{amsmath,amssymb,cite}
\usepackage[dvips]{graphicx}
\numberwithin{equation}{section}
\newcommand{\Cb}{{\mathbb C}}

\newcommand{\Zb}{{\mathbb Z}}
\newcommand{\Rb}{{\mathbb R}}
\newcommand{\J}{J}
\newcommand{\Jh}{\hat{\J}}
\newcommand{\JR}{\J^{(R)}}
\newcommand{\JF}{\J^{(F)}}
\newcommand{\sflow}[1]{{}^{\langle #1 \rangle}}
\newcommand{\slr}{{SL(2,\,$\Rb$)}{}}
\newcommand{\kh}{\hat{k}}
\newcommand{\del}{\partial}

\newcommand{\lz}{\ell}
\newcommand{\mz}{m}
\newcommand{\sz}{s}
\newcommand{\vl}{\vec{\ell}}
\newcommand{\vm}{\vec{m}}
\newcommand{\vs}{\vec{s}}
\newcommand{\vL}{\vec{L}}
\newcommand{\vM}{\vec{M}}
\newcommand{\vS}{\vec{S}}

\newcommand{\Lt}{\widetilde{L}}
\newcommand{\Mt}{\widetilde{M}}
\newcommand{\St}{\widetilde{S}}
\newcommand{\vb}{\vec{\beta}}
\newcommand{\chic}{\check{\chi}}
\newcommand{\cc}{\check{c}}

\newcommand{\Nc}{\check{N}}
\newcommand{\Th}{\Theta}
\newcommand{\Ic}{\check{I}}
\newcommand{\lc}{\check{\ell}}
\newcommand{\mC}{\check{m}}
\newcommand{\sC}{\check{s}}
\newcommand{\Lc}{\check{L}}
\newcommand{\Mc}{\check{M}}
\newcommand{\Sc}{\check{S}}
\newcommand{\Lct}{\widetilde{\check{L}}}
\newcommand{\Mct}{\widetilde{\check{M}}}
\newcommand{\Sct}{\widetilde{\check{S}}}
\newcommand{\jc}{\check{\jmath}}
\newcommand{\rc}{\check{r}}

\newcommand{\taub}{\bar{\tau}}

\DeclareMathOperator*{\Tr}{{\rm Tr}}
\DeclareMathOperator{\lcm}{{\rm lcm}}
\newcommand{\e}[1]{\,{\bf e}\!\left[#1\right]}
\renewcommand{\a}{{\sf a}}
\newcommand{\Ncal}{{\cal N}}
\newcommand{\NTwo}{$\Ncal=2$}
\newcommand{\Hcal}{{\cal H}}
\newcommand{\Hc}{\check{\Hcal}}

\newcommand{\deltam}[2]{\delta^{{\rm mod}\;#2}_{#1}}

\newcommand{\nn}{\nonumber}

\newcommand{\ra}{\rangle}
\newcommand{\la}{\langle}

\newcommand{\slKS}[1]{\frac{{\rm SL}(2,\,{\mathbb R})_{#1}}{\rm U(1)}}
\newcommand{\suKS}[1]{\frac{{\rm SU}(2)_{#1}}{\rm U(1)}}

\renewcommand{\a}{{\sf a}}

\newcommand{\Cc}{{\cal C}}
\newcommand{\OneAuthor}[3]{
\begin{center}
     \vspace{5mm}

    \noindent
    {\Large #1}

    \vspace{3mm}
    \noindent
    \hspace{0.7cm}{\it #2

    \vspace*{3mm}
    E-mail: {\tt #3}
}\end{center}
}
\newcommand{\KUCPlogo}{
\unitlength 1mm
\begin{picture}(30,10)(0,0)
 \put(0,0){{\Huge\boldmath$\cal K$}
              \kern -1.2em\raise 1ex\hbox{\large\sf UC}
              \kern -0.7em\hbox{\Huge\boldmath$\wp$}}
 \put(0,-2){\tiny\sl preprint}
\end{picture}
}

\newcommand{\preprintnumber}{KUCP-0178\\ 
 {\tt hep-th/0102176}}

\begin{document}
%

\thispagestyle{empty}
\KUCPlogo
\hfill
 February, 2001
\hfill
\hbox{\parbox[t]{3cm}{
\preprintnumber}}
%
%
%
%

\vspace{16mm}\noindent
\begin{center}
 {\LARGE
 Noncompact Gepner Models
 with Discrete Spectra
 }
\end{center}

\OneAuthor{Satoshi Yamaguchi}{Graduate School of Human and Environmental
          Studies, Kyoto University, Yoshida-Nihonmatsu-cho,
          Sakyo-ku, Kyoto 606-8501, Japan.}{yamaguch@phys.h.kyoto-u.ac.jp}

\vspace{0.2cm}\noindent
\rule{\textwidth}{0.4pt}

\noindent
{\bf Abstract\ \ }

{\small We investigate a noncompact Gepner model, which is composed of a
number of \slr/U(1) Kazama-Suzuki models and {\NTwo} minimal models. The
\slr/U(1) Kazama-Suzuki model contains the discrete series among the
{\slr} unitary representations as well as the continuous series. We
claim that the discrete series contain the vanishing cohomology and the
vanishing cycles of the associated noncompact Calabi-Yau manifold. We
calculate the Elliptic genus and the open string Witten indices.  In the
A${}_{N-1}$ ALE models, they actually agree with the vanishing
cohomology and the intersection form of the vanishing cycles.}


\vspace*{0.3cm}


\noindent
\rule{\textwidth}{0.4pt}

%
%
\baselineskip 3.5ex

\section{Introduction}
Recently, the worldsheet CFT of a noncompact singular Calabi-Yau
manifold has been largely investigated in the context of the holographic
description of the little string theories
\cite{GKP9907,GK9909,GK9911}. The modular invariant partition functions
have been constructed \cite{ES0002,Miz0003,Yam0007,Miz0009,NN0010}, and
D-branes on a noncompact singular Calabi-Yau manifold have been studied
by using the CFT descriptions\cite{SY0011,ES0011}.

However, the modular invariant partition function constructed in
\cite{ES0002,Miz0003,Yam0007,Miz0009,NN0010} includes only ``the
principal continuous series'' of the {\slr} unitary representations.  If
we naively treat this partition function\cite{Yam0007,SY0011}, the
elliptic genus in the closed string theory and the open string Witten
indices vanish, and we cannot obtain the vanishing cohomology of the
noncompact Calabi-Yau manifold, nor the intersection form of the
vanishing cycles of the manifold.  \footnote{In \cite{ES0011},
reasonable nonzero open string Witten indices are obtained by treating
the Liouville potential appropriately and using the Ramond ground states
of discrete series} Then, where are these compact cohomology and
vanishing cycles in the CFT description ?

Besides the principal continuous series, we can include the highest and
lowest weight discrete series\cite{DPL89}. We claim that we {\em should}
include this discrete series to obtain the right Calabi-Yau sigma model;
the vanishing cohomology and boundary states of the vanishing cycles
belong to this sector.

The wave functions of most of the states in the discrete series (we use
only these states) are ${\mathbb L}^2$ normalizable in the {\slr} group
manifold.  Therefore, they are localized near the deformed singularity
in the noncompact Calabi-Yau manifold, and they are ``bound states'' in
this theory. It is also reasonable that the compact cohomology belongs
to this class of states.

It is difficult, however, to construct a modular invariant partition
function of the {\slr} discrete series\cite{KS0001}. We use, in this
paper, the one proposed in \cite{HHRS91,HR9211,MO0001}. It includes the
new sectors obtained by ``spectral flow'' transformation --- a
series of automorphisms of the affine {\slr}.

We construct modular invariant partition functions of rather general
``noncompact Gepner models'', which consist of a number of \slr/U(1)
Kazama-Suzuki models with integral levels and {\NTwo} minimal models.
We also calculate the elliptic genera in these models in order to study
the topological properties and compare with the geometric ones. We also
construct the boundary states and calculate the open string Witten
indices between them in order to compare with the geometrical cycles and
the intersection form. We obtain nonzero closed and open string Witten
indices independent of the moduli parameter of the torus or annulus.

As an example, we consider the A$_{N-1}$ ALE model; this model consists
of an level $N$ \slr/U(1) Kazama-Suzuki model and an level $(N-2)$
{\NTwo} minimal model\cite{OV9511}.  As a result, we obtain the closed
string Witten index as $(N-1)$. This corresponds to the $(N-1)$
dimensional $(2,2)$ compact cohomology of the A$_{N-1}$ ALE space.  We
also obtain the open string Witten indices. They agree with the
intersection form of vanishing cycles in the A$_{N-1}$ ALE space.  This
intersection form also coincides with the one proposed in \cite{Ler0006}.

\section{{\slr/U(1)} Kazama-Suzuki model}
In this section, we construct the characters of the {\slr/U(1)}
Kazama-Suzuki model by summing up the new sectors
\cite{HHRS91,HR9211,MO0001}. This character is used to construct
``noncompact Gepner models'' in the next section.

In this paper, we use the convention in appendix A of \cite{SY0011}.

\subsection{Currents of {\slr/U(1)} Kazama-Suzuki model}

Let us first introduce the supersymmetric level $k$ {\slr} WZW model;
this consists of a set of level $\kh=k+2$ bosonic {\slr} currents
$\Jh^{\pm,3}$ and three free fermions $\psi^{\pm,3}$. These currents
satisfy the following OPE relations
\begin{gather*}
 \psi^3(z)\psi^3(w)\sim \frac{-1}{z-w},\quad
 \psi^+(z)\psi^-(w)\sim \frac{2}{z-w}, \\
 \Jh^3(z) \Jh^3(w)\sim \frac{-\kh/2}{(z-w)^2}, \quad
 \Jh^3(z) \Jh^{\pm}(w)\sim \frac{\pm \Jh^{\pm}(w)}{z-w},\quad
 \Jh^+(z)\Jh^-(w)\sim \frac{\kh}{(z-w)^2}+\frac{-2\Jh^{3}(w)}{z-w}.
\end{gather*}

We want to construct the \slr/U(1) Kazama-Suzuki model from these
currents. The supersymmetric affine U(1) in the denominator is generated
by the time-like boson $\J^3=\Jh^3+\frac12\psi^+\psi^-$ and the
time-like fermion $\psi^3$.  The {\NTwo} superconformal currents
$T(z),G^{\pm}(z),\JR(z)$ of the
Kazama-Suzuki model\cite{KS89} can be written as
\begin{align}
 T(z)&=\frac1{2k}\left(\Jh^+\Jh^-+\Jh^+\Jh^-\right)+\frac{\kh}{4k}
\left(\psi^+\del\psi^- +
 \psi^-\del\psi^+\right)+\frac1k\Jh^3\psi^+\psi^-,\qquad
 \frac c3=\frac{\kh}{k},\nn\\
 G^{\pm}(z)&=\frac1{\sqrt k}{\psi^{\mp}\Jh^{\pm}},\qquad
 \JR(z)=-\frac12 \psi^+\psi^- -\frac2k\J^3
    =-\frac2k\Jh^3-\frac{\kh}{2k}\psi^+\psi^-.\label{Currents}
\end{align}
Note that the $\psi^3$ decouples in these formulas. Therefore, we have
only to consider the decomposition
\begin{align*} 
(\text{bosonic }\widehat{\text{\slr}}_{\kh})\times(2 \text{ fermions }
 \psi^{\pm})\longrightarrow
 (\text{\slr/U(1) KS})\times(\widehat{\rm
 U(1)}_{-k}\ \J^3).
\end{align*}
If we denote the Verma module of $\widehat{\text{\slr}}_{\kh}$, two
fermions, and $\widehat{{\rm U}(1)}_{k}$ as $\Hc_{\lz},\Hcal_{\sz}$, and
$\Hcal_{\mz}$ respectively, the Verma module of the \slr/U(1)
Kazama-Suzuki model $\Hc_{\mz}^{\lz,\sz}$ is defined by the decomposition
\begin{align*}
 \Hc_{\lz}\otimes \Hcal_{\sz}=\sum_{\mz}\Hc_{\mz}^{\lz,\sz}
\otimes \Hcal_{\mz}.
\end{align*}

If we know the structures of $\Hc_{\lz},\Hcal_{\sz}$, and $\Hcal_{\mz}$,
then we can determine the structure of $\Hc_{\mz}^{\lz,\sz}$.  The Verma
modules of free fermions and $\widehat{\rm U(1)}_{-k}$ are not
difficult. We analyze them in subsection \ref{CosetCharSection}. In the
next subsection, we concentrate on the $\widehat{\text{\slr}}_{\kh}$,
especially, the character of the discrete series.

\subsection{$\widehat{\text{\slr}}_{\kh}$ discrete series}
Let us consider the lowest weight discrete representations of
 $\widehat{\text{\slr}}_{\kh}$.  Each of the Verma modules of ${\mathbb
 L}^2$ normalizable lowest weight representations of
 $\widehat{\text{\slr}}_{\kh}$ is labeled by a number $\lz \in \Zb,\
 1<\lz<\kh-1$.  \footnote{ The notation $j=\lz/2$ is often used in other
 papers. The Casimir eigenvalue of a representation can be written as
 $-j(j-1)$ by using this $j$.  } The character of this Verma module is
 known to be (for example, see \cite{HHRS91})
\begin{align*}
 \chi^+_{\lz}(\tau,z):=\Tr\left[q^{L_0-\frac{\kh}{8k}}y^{\Jh^3_0}\right]
         =\frac{q^{-\frac1{4k}\left(\ell-1\right)^2}y^{\frac12(\lz-1)}}{
         -i\theta_1(\tau,z)},
\end{align*}
where $q=\exp(2\pi i \tau),\ y=\exp(2\pi i z)$. This character is
divergent if we set $z=0$ because the representations of the zero modes
are infinite dimensional.

The modular transformation law of this character is not good. To make
this good, let us introduce a series of automorphisms of the algebra,
called ``spectral flow''.
\begin{align*}
 &\Jh^3_n\sflow{v}=\Jh^3_n-\frac{\kh}{2}\delta_{n,0},\quad
 \Jh^{\pm}_n\sflow{v}=\Jh^{\pm}_{n\pm v}, \quad v\in \Zb.
\end{align*}
These new generators $\Jh^{\pm,3}_n\sflow{v}$ satisfy the same algebra
as $\Jh^{\pm,3}_n$.

The character of the new sector obtained by this spectral flow operation
can be written as
\begin{align*}
  \chi^+_{\lz}\sflow{v}(\tau,z)&:=\Tr\left[q^{L_0\sflow{v}
-\frac{\kh}{8k}}y^{\Jh^3\sflow{v}_0}\right]\\
&=\frac{(-1)^vq^{-\frac1{4k}\left(\lz-1-kv\right)^2}
y^{-\frac1{2k}\left(\lz-1-kv\right)}}{-i\theta_1(\tau,z)}.
\end{align*}

Now, we sum up for $v \in 2\Zb$ in order to make a character with good modular
properties
\begin{align*}
 \sum_{v\in
 2\Zb}\chi^{+}_{\lz}\sflow{v}(\tau,z)=\frac{\Th_{\lz-1,-k}(\tau,z)}%
{-i\theta_1(\tau,z)}.
\end{align*}
This character includes a negative level theta function. Therefore, this
character diverges in the region ${\rm Im} \tau
>0$. Here, we treat this character as a formal series.

If we sum up for odd integer $v$, we obtain the same result as the case
of the $(\kh-\lz)$ highest weight representation for even $v$
\begin{align*}
 \sum_{v\in 2\Zb+1}\chi^{+}_{\lz}\sflow{v}(\tau,z)
=\sum_{v\in 2\Zb}\chi^{-}_{\kh-\lz}\sflow{v}(\tau,z)
=\frac{-\Th_{-k+\ell-1,-k}(\tau,z)}%
{-i\theta_1(\tau,z)}.
\end{align*}
This is because the $v=1$ spectral flow maps $\lz$ lowest weight
representation to $(\kh-\lz)$ highest weight representation \cite{MO0001}.
We have only to consider the lowest weight representations for this
reason.

Next, we define the following character $\chic_{\lz}$, and use this character
in the rest of this paper
\begin{align}
 \chic_{\lz}(\tau,z)&:=\sum_{v\in 2\Zb}\chi^{+}_{\lz}\sflow{v}(\tau,z)+
\sum_{v\in 2\Zb+1}\chi^{+}_{\kh-\lz}\sflow{v}(\tau,z)
=\frac{\Th_{\lz-1,-k}(\tau,z)-\Th_{-(\lz-1),-k}(\tau,z)}{-i\theta_1(\tau,z)}.
\label{SL2RChar}
\end{align}
Note that each coefficient of the power of $q$ in this character is
convergent even if we set $z=0$ , in contrast with $\chi^{+}_{\lz}$ or
$\sum_{v\in 2\Zb}\chi^{+}_{\lz}\sflow{v}$ . But the sum is still
divergent.

\subsection{The characters of \slr/U(1)}
\label{CosetCharSection}
The characters of the $\widehat{{\rm U}(1)_{-k}}$ generated
by $\J^3=\Jh^3+\frac12\psi^+\psi^-$ is expressed as
\begin{align}
\Tr_{\Hcal_{\mz}}[q^{L_0-\frac{1}{24}}y^{\J^3_0}]
=\Theta_{\mz,-k}(\tau,z)/\eta(\tau),\quad \mz\in \Zb_{2k}.
\label{U1Char}
\end{align}
This formula includes a theta function of negative level and
is divergent because the $J^3$ direction is time-like. We treat
this character as a formal series as the same way as $\chic_{\lz}$.

There remaining is the characters of two fermions
($\widehat{{\rm SO(2)}}_1$). This character can be written as
\begin{align}
 \Tr_{\Hcal_{\sz}}[q^{L_0-\frac{1}{24}}y^{\JF_0}]
=\Theta_{\sz,2}(\tau,z)/\eta(\tau),\quad \sz\in \Zb_4,
\label{FermionChar}
\end{align}
where $\JF_0$ is the zero mode of the fermion number current
$\JF=\frac12\psi^+\psi^-$ .

Collecting the characters (\ref{SL2RChar}), (\ref{U1Char}),
(\ref{FermionChar}) , and looking at the forms of the currents
(\ref{Currents}), we obtain the characters of \slr/U(1) KS model
$\chic^{\lz,\sz}_{\mz}:=
\Tr_{\Hc^{\lz,\sz}_{\mz}}[q^{L_0-\frac{c}{24}}y^{\JR_0}]$ through 
the decomposition
\begin{align}
 \chic_{\lz}\left(\tau,-\frac2k z_1+z_2\right)
\Th_{\sz,2}\left(\tau,-\frac{\kh}{k}z_1+z_2\right)
=\sum_{m_0\in\Zb_k}\chic^{\lz,\sz}_{\mz}(\tau,z_1)
\Th_{\mz,-k}\left(\tau,z_2\right).\label{KSChar}
\end{align}
In this formula, $\chic_{\lz}$ is a divergent series, however,
$\Th_{\mz,-k}\left(\tau,z_2\right)$ is also divergent. Therefore, the
character $\chic^{\lz,\sz}_{\mz}(\tau,z)$ is possibly convergent.

Let us write the explicit form of $\chic^{\lz,\sz}_{\mz}(\tau,z)$.  For
 this purpose, we define the ``string function'' $\cc^{\lz}_{m'}$ of
 {\slr} as
\begin{align*}
 \chic_{\lz}(\tau,z)=\sum_{m'\in
 \Zb_{\kh}}\cc^{\lz}_{m'}(\tau)\Th_{m',-\kh}(\tau,z).
\end{align*}
Using this string function, we express the character of \slr/U(1)
Kazama-Suzuki model as
\begin{align*}
 \chic^{\lz,\sz}_{\mz}(\tau,z)=\sum_{r\in \Zb_{\kh}}\cc^{\lz}_{m-s+4r}(\tau)
\Th_{-2m-k(s+4r),2\kh k}(\tau,z/k).
\end{align*}
Note that if all the string functions $\cc^{\lz}_{m-s+4r}(\tau)$ are
convergent, then $\chic^{\lz,\sz}_{\mz}(\tau,z)$'s are convergent.
The explicit form of $\cc^{\lz}_{m'}(\tau)$ can be obtained by using the
results of \cite{Sfe91}. If $\lz+m'$ is an odd integer, 
then $\cc^{\lz}_{m'}=0$. If $\lz+m'$ is an even integer, then
\begin{align*}
 \cc^{\lz}_{m'}=\eta(\tau)^{-3}\sum_{r\in \Zb}\sum_{u=0}^{\infty}
(-1)^u\left[
q^{-k\left(\frac{\lz-1}{2k}+\frac{u}{2}\right)^2
   +\kh\left(\frac{m'}{2\kh}+\frac{u}{2}+r\right)^2}
+
q^{-k\left(\frac{\lz-1}{2k}-\frac{u+1}{2}\right)^2
   +\kh\left(\frac{m'}{2\kh}+\frac{u+1}{2}+r\right)^2}
\right].
\end{align*}
This function is shown to be actually convergent in
\cite{HR9211}. Consequently, we conclude that the character
$\chic^{\lz,\sz}_{\mz}(\tau,z)$ is convergent.

Let us consider the modular transformation properties of these
characters in order to construct modular invariant partition
functions. The modular transformation law can be read from
(\ref{KSChar}) as follows.
\begin{align*}
& \chic^{\lz,\sz}_{\mz}(\tau+1,z)=
\e{-\frac{\lz(\lz-2)}{4k}+\frac{\sz^2}{8}
+\frac{\mz^2}{4k}-\frac{\kh}{8k}}\chic^{\lz,\sz}_{\mz}(\tau,z),\\
&\chic^{\lz,\sz}_{\mz}(-1/\tau,z/\tau)=\e{\frac{\kh}{2k}
\frac{z^2}{\tau}}
\sum_{\lz,\mz,\sz}(-1)\sqrt{\frac 2k}\sin \pi
 \frac{(\lz-1)(\lz'-1)}{k}\\
&\hspace{7cm}\times\frac{1}{\sqrt{8k}}
\e{-\frac{\sz\sz'}{4}-\frac{\mz\mz'}{2k}}\chic^{\lz',\sz'}_{\mz'}(\tau,z).
\end{align*}

Note that from (\ref{KSChar}), we can read off the charge of the
states in $\Hc^{\lz,\sz}_{\mz}$ to be $-\sz/2-\mz/k \mod 2$. This relation
is used to perform the GSO projection.

Also, it is convenient to define a ``index''
$\Ic^{\lz}_{\mz}(\tau,z):=\chic^{\lz,1}_{\mz}(\tau,z)
-\chic^{\lz,-1}_{\mz}(\tau,z)$ in order to construct the elliptic genera
and Witten indices.  This $\Ic^{\lz}_{\mz}$ has the following
properties from (\ref{KSChar}).\label{sl2u1char}
\begin{align}
\Ic^{\lz}_{\mz}(\tau,0)=-\delta_{\mz,\lz-1}+\delta_{\mz,-(\lz-1)}.
\label{Index}
\end{align}
$\Ic^{\lz}_{\mz}(\tau,0)$ actually has no $\tau$ dependence because of
worldsheet supersymmetry; there are only contributions from Ramond
ground states.

\section{Noncompact Gepner models}

In this section, we construct the ``noncompact Gepner models''
constructed by a number of integral level \slr/U(1) Kazama-Suzuki models
and {\NTwo} minimal models (SU(2)/U(1) Kazama-Suzuki models).  The
Gepner-like description of ALE\cite{OV9511}, and Seiberg-Witten
curve\cite{Ler0006} are included in this class of models.

In this paper, we construct only the discrete part. However, there is
also the continuous part for each \slr/U(1).

\subsection{The closed string theory}

Let us first consider the closed string theory and construct the
toroidal partition function. The theory we consider here is
\begin{align*}
\left(\slKS{\Nc_1}\times \slKS{\Nc_2}\times \dots 
\times \slKS{\Nc_{\rc}}\times
\suKS{N_1}\times \suKS{N_2}\times
 \dots \times \suKS{N_{r}}\right)/\Zb_K, 
\end{align*}
where $K=\lcm(\Nc_{\jc},N_j)|_{\jc=1,\dots,\rc,\ j=1,\dots,r}$. This
orbifold projection is the one onto the integer charged states. 
This theory can be expressed as an {\NTwo} Landau-Ginzburg orbifold with
a superpotential
\begin{align*}
 W=Y_1^{-\Nc_1}+Y_2^{-\Nc_2}+\dots+Y_{\rc}^{-\Nc_{\rc}}
+X_{1}^{N_1}+\dots+X_{r}^{N_r},
\end{align*}
where $Y_1,\dots,Y_{\rc},X_1,\dots,X_{r}$ are chiral superfields.
We construct the partition function of this theory by the beta-method
\cite{Gep88} and determine the combination of left mover and right
mover.

The central charge of this theory is expressed as
\begin{align*}
c/3&=\sum_{\jc=1}^{\rc}\frac{\Nc_{\jc}+2}{\Nc_{\jc}}
+\sum_{j=1}^{r}\frac{N_{j}-2}{N_{j}}
=r+\rc+\sum_{\jc=1}^{\rc}\frac{2}{\Nc_{\jc}}
-\sum_{j=1}^{r}\frac{2}{N_{j}}.
\end{align*}
Since we expect that this theory is equivalent to a sigma model on a
Calabi-Yau manifold, we consider the case that $c/3$ is an integer. We
also concentrate to the case that $\rc+r-c/3$ is an even integer for a
technical reason.

A Verma module of this theory is a tensor product of ones of each
sub-theory.  The character of the Verma module $f_{\a}(\tau)$ is a
product of ones of each sub-theory
\begin{align*}
 f_{\a}(\tau,z):=\chic^{\lc_1,\sC_1}_{\mC_1}(\tau,z)
\dots\chic^{\lc_{\rc},\sC_{\rc}}_{\mC_{\rc}}(\tau,z)
\chi^{\ell_{1},s_{1}}_{m_{1}}(\tau,z)
\dots \chi^{\ell_{r},s_{r}}_{m_{r}}(\tau,z),
\end{align*}
where $\chi^{\ell,s}_m$ is an {\NTwo} minimal model character\cite{Gep88},
 and the label $\a$ is defined as
\begin{align*}
 &\vl=(\lc_1,\dots,\lc_{\rc};\ell_1,\dots,\ell_r),
 \quad \vm=(\mC_1,\dots,\mC_{\rc};m_1,\dots,m_r),
 \quad \vs=(\sC_1,\dots,\sC_{\rc};s_1,\dots,s_r),\\
 & \a=(\vl,\vm,\vs),
\end{align*}
and each $(\lc_j,\mC_j,\sC_j)$ is a label of a Verma module of
\slr${}_{\Nc_j}$/U(1), and each $(\ell_j,m_j,s_j)$ is that of
SU(2)${}_{N_j}$/U(1).  We define, for the sake of convenience, the inner
products between two $\vm$'s, and between two $\vs$'s as follows.
\begin{align*}
& \vm\bullet\vm':=-\sum_{\jc}\frac{\mC_{\jc}\mC'_{\jc}}{2\Nc_{\jc}}
+\sum_{j}\frac{m_{j}m'_{j}}{2N_{j}},\quad
 \vs\bullet\vs':=-\sum_{\jc}\frac{\sC_{\jc}\sC'_{\jc}}{4}
-\sum_{j}\frac{s_{j}s'_{j}}{4}.
\end{align*}
Also, we introduce a special vector $\vb=(2,\dots,2;2,\dots,2)$ which is
the same type of $\vm$. Using this vector, we can write the charge
integrality condition as $\vb\bullet\vm\in \Zb$.

Next, we need the modular transformation laws of $f_{\a}$ to construct the
modular invariant partition function and determine the combination of
left mover and right mover. These modular properties can be written as
\begin{align}
 & f_{\a}(\tau+1,z)=\e{
-\sum_{\jc}\frac{\lc_{\jc}(\lc_{\jc}-2)}{4\Nc_{\jc}}
+\sum_{j}\frac{\ell_{j}(\ell_{j}+2)}{4N_{j}}
-\frac12(\vs \bullet \vs+\vm \bullet \vm)
+\frac{c}{12}}f_{\a}(\tau,z),\nn\\
& f_{\a}(-1/\tau,z/\tau)=\e{\frac{c}{6}
\frac{z^2}{\tau}}\sum_{\a'}S_{\a\a'}
f_{\a'}(\tau,z),\nn\\
&S_{\a\a'}:=\left(\prod_{\jc=1}^{\rc}(-1)A_{\lc_{\jc}-2,\lc_{\jc}'-2}\right)
\left(\prod_{j=1}^{r} A_{\ell_j\ell_j'}\right)
\frac{1}{8N}
\e{\vs\bullet\vs'+\vm\bullet\vm'},
\label{ModularProperty}
\end{align}
where $A_{\ell\ell'}$'s are $SU(2)$ S matrices.

With these notations, we can write the modular invariant partition
function of NS sector as
\begin{align*}
 Z(\tau,\taub)=\frac{1}{2^{\rc+r}}
\sum_{b=0}^{K-1}
\sum_{\substack{\vl,\vm,\vs,\bar{\vs} \\ 
\sC_{\jc},s_j,\bar{\sC}_{\jc},\bar{s}_j,=0,2\\
 \vb\bullet\vm\in \Zb}}
f_{(\vl,\vm,\vs)}(\tau)
\bar f_{(\vl,\vm+b\vb,\bar{\vs})}(\taub).
\end{align*}
We can check that this partition function is actually invariant under
the transformation (\ref{ModularProperty}).

Now, let us construct the elliptic genus in order to investigate the
topological feature of this model.  The elliptic genus
$I(\tau,\taub,z):= \Tr_{\rm RR}[(-1)^{F}q^{L_0-c/24}\bar q^{\bar
L_0-c/24}y^{J_0}]$ becomes a sum of products
\begin{align}
& I(\tau,\taub,z)=\frac{1}{2^{\rc+r}}\sum_{b=0}^{K-1}
\sum_{\substack{\vl,\vm \\
\vb\bullet\vm\in \Zb}}I^{\vl}_{\vm}(\tau,z)\bar I^{\vl}_{\vm+b\vb}(\taub,0),\nn\\
& I^{\vl}_{\vm}(\tau,z)=\Ic^{\lc_{1}}_{\mC_1}(\tau,z)
\dots\Ic^{\lc_{\rc}}_{\mC_{\rc}}(\tau,z)I^{\ell_{1}}_{m_{1}}(\tau,z)
\dots I^{\ell_{r}}_{m_{r}}(\tau,z),
\label{EllipticGenus}
\end{align}
where $I^{\ell}_{m}=\chi^{\ell,1}_{m}-\chi^{\ell,-1}_{m}$ and
$\Ic^{\lc}_{\mC}$ is defined in section \ref{sl2u1char}.  We can check
that this elliptic genus has the right modular properties\cite{KYY9306}.

The Witten index is obtained as $I(\tau,\taub,0)$. This is actually
independent of $\tau$ from the relation (\ref{Index}).

\subsection{D-branes}
In this subsection, we consider the boundary states, the annulus
 amplitude and the open string Witten indices in the model defined in
 the previous subsection. We use almost the same method as
 \cite{BDLR9906}, and we only show the results here.

There are two types of gluing conditions of the {\NTwo} superconformal
algebra: the A-type and the B-type\cite{OOY9606}. In both cases, the boundary
conditions (boundary states) are labeled by $\alpha=(\vL,\vM,\vS)$ which
is the same label as $\a$.

In the case of the A-type gluing condition, the open string partition
function becomes
\begin{align*}
Z^{\rm A}_{\alpha\tilde\alpha}&:=
\Tr_{\alpha\tilde \alpha,\rm A, NS}\left[q^{L_0-c/24}\right]\nn\\
&=\frac{1}{2^{\rc+r}}
\sum_{\a'}^{\rm NS}\sum_{b=0}^{K-1}
\left(\prod_{\jc=1}^{\rc}N^{\lc_{\jc}'-2}_{\Lc_{\jc}-2,\Lct_{\jc}-2}
\deltam{-\Mc_{\jc}+\Mct_{\jc}+\mC'_{\jc}+2b}{2\Nc_{\jc}}\right)
\left(\prod_{j=1}^{r}N^{\ell'_j}_{L_{j}\Lt_{j}}
\deltam{-M_{j}+\Mt_{j}+m'_{j}+2b}{2N_{j}}\right)f_{\a'}(\tau),
\end{align*}
where, $N^{\ell}_{L\tilde L}$'s are SU(2) fusion coefficients.

The A-type open string Witten index becomes
\begin{align}
I^{\rm A}_{\alpha\tilde\alpha}:=
\Tr_{\alpha\tilde \alpha,\rm A, R}\left[q^{L_0-c/24}(-1)^{F}\right]
=
(-1)^{\frac12(\St-S)}
\sum_{b=0}^{K-1}
\left(\prod_{\jc}N^{\Mc_{\jc}-\Mct_{\jc}+2b-2}_{\Lc_{\jc}-2,\Lct_{\jc}-2}\right)
\left(\prod_{j}N^{M_{j}-\Mt_{j}+2b-2}_{L_{j}\Lt_{j}}\right),
\label{AIntersection}
\end{align}
where $S=\sum_{\jc=1}^{\rc}\Sc_{\jc}+\sum_{j=1}^{r}S_j,\ 
\St=\sum_{\jc=1}^{\rc}\Sct_{\jc}+\sum_{j=1}^{r}\St_j$ .

On the other hand, in the B-type case, open string partition function
becomes
\begin{align*}
Z^{\rm B}_{\alpha\tilde\alpha}:=
\Tr_{\alpha\tilde \alpha,\rm B, NS}\left[q^{L_0-c/24}\right]
=\frac{1}{2^{\rc+r}}
\sum_{\a'}^{\rm NS}
\left(\prod_{\jc=1}^{\rc}N^{\lc_{\jc}'-2}_{\Lc_{\jc}-2,\Lct_{\jc}-2}\right)
\left(\prod_{j=1}^{r}N^{\ell'_j}_{L_{j}\Lt_{j}}\right)
\deltam{-M+\Mt+K\vb\bullet\vm}{K}
f_{\a'}(\tau),
\end{align*}
where, $M=K\vb\bullet\vM,\ \Mt=K\vb\bullet\vec{ \tilde M} $ . The open
string Witten index is written as
\begin{align}
I^{\rm B}_{\alpha\tilde\alpha}:=
\Tr_{\alpha\tilde \alpha,\rm B, R}\left[q^{L_0-c/24}(-1)^{F}\right]
&=
(-1)^{\frac12(\St-S)}\sum_{\vm'}
\left(\prod_{\jc=1}^{\rc}N^{\mC_{\jc}'-1}_{\Lc_{\jc}-2,\Lct_{\jc}-2}\right)
\left(\prod_{j=1}^{r}N^{m'_j-1}_{L_{j}\Lt_{j}}\right)
\deltam{-M+\Mt+K\vb\bullet(\vm'+1/2 \vb)}{K}.
\label{BIntersection}
\end{align}

In both types of gluing conditions, the open string Witten indices are
actually independent of $\tau$.

\subsection{An Example --- ALE space ---}
In this subsection, we consider the properties of the A${}_{N-1}$ ALE
model in detail; this model is composed of an \slr${}_N$/U(1)
Kazama-Suzuki model and an {\NTwo} level $(N-2)$ minimal
model\cite{OV9511}.

The elliptic genus of the A${}_{N-1}$ ALE model is obtained as the
special case of (\ref{EllipticGenus}). In particular, the closed string
Witten index becomes $ I(\tau,\taub,z=0)=N-1$ .  This exactly correspond
to the $(N-1)$ dimensional (2,2) compact cohomology elements of the
A${}_{N-1}$ ALE space. Other compact cohomology elements of the
A${}_{N-1}$ ALE space are known to be 0 dimensional.

Associated open string Witten indices are also obtained as the special
case of (\ref{AIntersection}) and (\ref{BIntersection}). In this special
case, the A-type open string Witten indices are the same as the B-type
ones and the result is
\begin{align}
 I_{\alpha\tilde\alpha}=(-1)^{(S-\St)/2}\sum_{m=0}^{2N-1}N_{\Lc_1-2,\Lct_1-2}^{M-\Mt+m}
N_{L_1\Lt_1}^{m}. \label{ALEResult}
\end{align}
This fact the A-type open string Witten indices coincides with the
B-type ones is consistent to the fact that the A${}_{N-1}$ ALE space
is self mirror.

If we set $\Lc_1=\Lct_1=2$ in (\ref{ALEResult}), 
this gives the same result as \cite{Ler0006}
and it is the correct intersection form of the vanishing 2-cycles of
the A${}_{N-1}$ ALE space. In particular, this intersection form becomes
the A${}_{N-1}$ extended Cartan matrix when we set $S=\St=L_1=\Lt_1=0$.

\begin{figure}[t]
 \begin{center}
  \includegraphics[width=4.5cm]{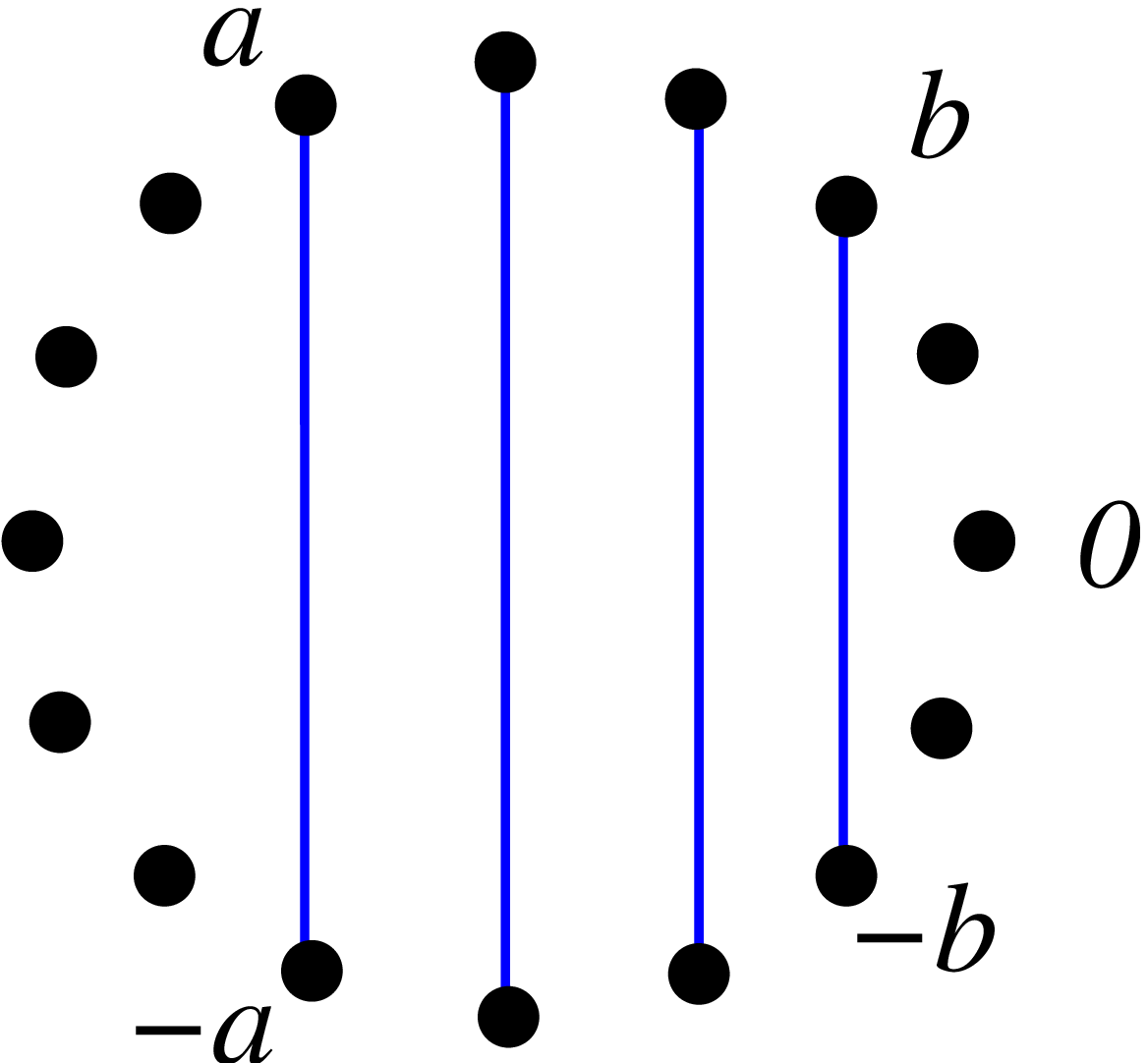}
 \end{center}
 \caption{The image of the cycle associated with a boundary state
of $\Lc_1,L_1,M=-1,S=0$ with $\Lc_1+L_1\in 2\Zb$. The total space is
the $x$-plane when we express the ALE space as $x^N+y^2+z^2=\mu$ in $\Cb^3$.
the dots are the root of the equation $x^N=\mu$. A line between two dots is a
 2-cycle $\Cc_{-\nu,\nu},\ \nu=b,b+1,\dots,a,\ a=(\Lc_1+L_1-1)/2, \ b=|\Lc_1-L_1-2|/2+1/2$. The sum of these cycles
 corresponds to the boundary state mentioned above.}
 \label{Cycles}
\end{figure}

In the general values of $\Lc_1$ and $L_1$ , the associated cycles in the
ALE space is as follows. Let us denote the 2-cycles in ALE space as
$\Cc_{\nu},\ \nu\in \Zb,\ \Cc_{\nu+N}=\Cc_{\nu}
,\ \sum_{\nu=0}^{N-1}\Cc_{\nu}=0$. Here, we set the intersection form as
$\la\Cc_{\nu},\Cc_{\nu'}\ra=2\deltam{\nu,\nu'}{N}-\deltam{\nu,\nu'+1}{N}
-\deltam{\nu,\nu'-1}{N}$. 
We also define $\Cc_{\nu_1\nu_2}:=\Cc_{\nu_1}
+\Cc_{\nu_1+1}+\dots+\Cc_{\nu_2-1}$ for the sake of convenience.
For the boundary states with $\Lc_1\le (N+2)/2,\ L_1\le (N-2)/2,\
\Lc_1+L_2 \in 2\Zb, \  M=S=0$, we can write the associated cycle as
\begin{align*}
 &\gamma_{(\Lc_1,L_1,M=0,S=0)}=\Cc_{-a+1,a}+\Cc_{-a+2.a-1}+\dots+\Cc_{-b,b+1},
\\& \text{ where, } a=\frac12(\Lc_1+L_1), \ b=\frac12|\Lc_1-L_1-2|.
\end{align*}
On the other hand, for the boundary states with $\Lc_1\le (N+2)/2,\ L_1\le (N-2)/2,\
\Lc_1+L_2 \in 2\Zb+1, \  M=-1, \ S=0$, we can write the associated cycle as
\begin{align*}
 &\gamma_{(\Lc_1,L_1,M=-1,S=0)}=\Cc_{-a,a}+\Cc_{-a+1.a-1}+\dots+\Cc_{-b,b}
\\& \text{ where, } a=\frac12(\Lc_1+L_1-1), \ b=\frac12|\Lc_1-L_1-2|+\frac12.
\end{align*}
Actually, the open string Witten indices (\ref{ALEResult}) and the
intersection number between these cycles are the same. The image of this cycle is shown in
Fig.\ref{Cycles}. The cycles for general values of $M$ are obtained by
shifting by the $Z_N$ symmetry.

\section{Conclusion}
In this paper, we investigate the discrete series of the noncompact
Gepner model. We claim that the compact cohomology and the compact
cycles belong to this sector. We check it in the case of the A${}_{N-1}$
ALE model. We treat only the A-type ALE in this paper, but D-type and
E-type ALE can be also treated as the same way and the results will be
the same as \cite{LLS0006}.

Other interesting noncompact models are ADE type Calabi-Yau 3-fold and
4-fold singularities. In these cases, the level of the \slr/U(1) becomes
fractional and this model does not belong to the models treated in this
paper.  We should consider some multiple covering of the {\slr} group
manifold and a fractional $\ell$ in order to treat the discrete series
of the fractional level \slr/U(1) Kazama-Suzuki models. We postpone this
in a future work.

The relation between the continuous series and discrete series is also
interesting. At the level of the modular invariance, these two are
independent of each other. However, in the OPE level, they are
interacting.

\subsection*{Acknowledgement}
The author would like to thank Tsuneo Uematsu and Katsuyuki Sugiyama for
useful discussions and encouragement.  I am also grateful to Michihiro Naka,
Masatoshi Nozaki, Yuji Satoh, Yuji Sugawara, Toshiaki Tanaka, Kentaroh
Yoshida for useful discussions and comments.  I would also like to thank
the organizers (T.~Eguchi et al) of the Summer Institute 2000 at
Yamanashi, Japan, 7-21 August, 2000, where a part of this work is done.

This work is supported in part by the JSPS Research Fellowships
for Young Scientists.

\providecommand{\href}[2]{#2}\begingroup\raggedright\endgroup

\end{document}